\documentclass{optica-article}
\journal{opticajournal}
\articletype{Research Article}

\usepackage{lineno}
\usepackage{subcaption}
\usepackage{amsmath}
\usepackage{siunitx}
\usepackage{booktabs}

\newcommand{\e}{\mathrm{e}}
\newcommand{\ii}{\mathrm{i}}

\newcommand{\sech}{\mathrm{sech}}

\begin{document}
	\title{Frequency-Multiplexed Photonic Reservoir Computing with a Synchronously Pulse-Driven Optical Cavity}
	\author{Amir Arsalan Arabieh\authormark{1,2,*},
		Simon-Pierre Gorza\authormark{2},
		Serge Massar\authormark{1}}
	\address{\authormark{1}Laboratoire d’Information Quantique, CP 224, Université libre de Bruxelles (ULB), Av. F. D. Roosevelt 50, 1050 Brussels, Belgium\\
		\authormark{2}Service OPERA-Photonique, CP 194/5, Université libre de Bruxelles (ULB), Av. F. D. Roosevelt 50, 1050 Brussels, Belgium}
	\email{\authormark{*}amirarsalan.arabieh@ulb.be}

	\begin{abstract}
		In this work, we introduce a new platform for frequency-multiplexed reservoir computing based on an optical cavity driven by pulses whose repetition period is matched to the cavity roundtrip time. The input symbols are encoded by modulating either the amplitude or the phase of the driving pulses. Our numerical results show that the proposed synchronously pulse-driven cavity operates in a weakly nonlinear regime under anomalous dispersion, while under normal dispersion it exhibits optical bistability with a high-peak-power upper-branch state. By isolating the key physical parameters, we quantify their individual contributions to information-processing performance. We further show that spectral symmetry breaking induced by third-order dispersion and Raman scattering almost doubles the information-processing capacity. Finally, we demonstrate that modulation-induced branch switching within the bistable regime limits stable reservoir operation.
	\end{abstract}
	\section{Introduction}

		 The growing computational and energy demands of modern machine-learning systems motivate the development of alternative hardware platforms for efficient information processing~\cite{Maslej2025,Sevilla2022,Sze2017,Kachris2025}. Optical neural networks (ONNs) offer a potential alternative to their digital counterparts~\cite{Denz2013,Shastri2021,Genty2021}. Native optical nonlinearities have been exploited for information processing, including second-harmonic generation~\cite{Wright2022} and the nonlinear interactions of modes propagating through multimode fibers~\cite{Tegin2021}. In-memory linear transformations have been implemented using coherent meshes of Mach--Zehnder interferometers~\cite{Shen2017}, as well as photonic phase-change-material weight elements with integrated waveguide routing~\cite{Feldmann2021}.
		 
		 Photonic neuromorphic computing encompasses neural-inspired information-processing architectures implemented in photonic hardware~\cite{Christensen2022,Yan2024}. Among different machine-learning paradigms considered for neuromorphic photonics, extreme learning machines (ELMs)~\cite{Huang2004,Huang2006} and reservoir computing~\cite{Jaeger2004,Jaeger2009} have attracted considerable attention. They provide conceptually simple counterparts to feed-forward and recurrent neural architectures, respectively, while typically requiring only the output readout to be trained. This simplified training procedure makes both approaches particularly well suited to physical implementations.
		 
		 Among the physical mechanisms that can be exploited to implement such architectures, nonlinear wave propagation was theoretically established as a computational layer for ELMs in Ref.~\cite{Marcucci2020}, where dynamics governed by equations such as the nonlinear Schrödinger equation~\cite{Agrawal2012} were shown to provide rich feature mappings for learning. Subsequent studies have extensively investigated nonlinear and dispersive propagation in optical fibers as a physical substrate for ELMs~\cite{Zhou2022,Saeed2025}, whereas this aspect remains relatively underexplored for reservoir computing.
		 
		 In this context, a particularly compelling approach is frequency-multiplexed reservoir computing, in which distinct computational nodes are accessed through the frequency degree of freedom of light~\cite{Butschek2022}. Within this framework, recent studies have begun to exploit the nonlinear dynamics of optical resonators for reservoir computing. These include a numerical investigation of chaotic frequency-comb dynamics in Kerr microresonators for temporal information processing~\cite{Shishavan2025}, and image classification~\cite{Shishavan2026}, as well as experimental demonstrations of cavity solitons~\cite{Akhmediev2005,Akhmediev2008,Hasegawa1989, Leo2010} as computational resources in both microring resonators~\cite{Cuevas2025} and fiber cavities~\cite{arsalan2026}. These implementations primarily evaluate computational performance through specific benchmark tasks, including time-series prediction, nonlinear equalization, and image classification.
		 
		To the best of our knowledge, the present study is the first numerical investigation of a frequency-multiplexed reservoir computing platform based on a fiber cavity driven by picosecond optical pulses whose repetition period is matched to the cavity roundtrip time (see Fig.~\ref{fig:model}). Synchronously pulse-driven Kerr cavities have previously been investigated numerically in the context of dissipative Kerr-soliton generation~\cite{Malinowski2017} and demonstrated experimentally in a fiber-based Fabry--Pérot resonator~\cite{Obrzud2017}, where the continuous-wave (CW) pump was replaced by a periodic train of optical pulses. The present reservoir computing implementation builds upon our recent work in Ref.~\cite{arsalan2026}, where CW-driven cavity solitons served as a nonlinear computational substrate. The present architecture does not rely on the formation of dissipative cavity solitons or sustained chaotic dynamics, which allows us to operate over a broader range of parameters. We adopt a parameter set similar to that used in Ref.~\cite{arsalan2026} as a starting point and perform one-parameter sweeps to investigate the influence of distinct physical and operating parameters on reservoir performance.
		
		We use Information Processing Capacity (IPC)~\cite{dambre2012} as the primary metric to assess the performance of the proposed reservoir. Unlike conventional task-specific reservoir computing benchmarks, IPC decomposes the reservoir response into linear and higher-order nonlinear contributions, making it a useful diagnostic tool for unexplored physical systems. Using this metric, we investigate synchronous pulse pumping in both the anomalous- and normal-dispersion regimes. These pulse-driven states substantially differ from the CW-driven cavity-soliton state investigated in Ref.~\cite{arsalan2026}. In the anomalous-dispersion regime, we observe no optical bistability over the investigated parameter range; depending on the cavity detuning and driving strength, the system either exhibits chaotic dynamics or converges to a stationary state. By contrast, under normal dispersion, the cavity exhibits optical bistability with a high-peak-power upper-branch state. We also compare amplitude and phase encoding of the input symbols and quantify their distinct effects on the cavity dynamics, computational performance, and dynamical stability.
		
		The paper is organized as follows. In Sec.~\ref{sec:RC_computing}, we review the reservoir computing framework and IPC measures for dynamical systems. In Sec.~\ref{sec:PU_RC}, we present the synchronously pulse-driven optical cavity as an optical platform for frequency-multiplexed reservoir computing. In Sec.~\ref{sec:FOI}, we investigate the influence of the dominant physical parameters of the fiber cavity, including group-velocity dispersion, cavity detuning, input driving strength, and effective loss, to clarify their respective roles in information processing. In Sec.~\ref{sec:HOI}, we examine the effects of higher-order perturbations, including third-order dispersion and Raman scattering, compare phase and amplitude encoding, assess the robustness of the system in the presence of noise, and evaluate its performance on NARMA benchmark tasks. In Sec.~\ref{sec:bistable}, we investigate reservoir computing in the bistable regime by comparing three operating configurations. We revisit the CW-driven cavity-soliton state used for information processing in our previous work~\cite{arsalan2026} and compare it with the upper-branch solution of the synchronously pulse-driven cavity under normal dispersion, both corresponding to bistable operation, and with the synchronously pulse-driven cavity under anomalous dispersion, for which no bistability is observed over the investigated parameter range. Finally, Sec.~\ref{sec:conclusion} summarizes our findings and outlines directions for future research.

	\section{Reservoir Computing and Information Processing Capacity}
	\label{sec:RC_computing}
	\paragraph{Reservoir computing.}
	Reservoir computing (RC) is a recurrent neural network (RNN) framework for processing time series in which only the readout is trained, while the internal recurrent connections are fixed and typically randomized. At a discrete time step $m$, an RC system consists of (i) an input layer that injects the input $u(m)$ through input weights $\mathbf{W}_{\mathrm{in}}$, (ii) a reservoir with state vector $\mathbf{x}(m)\in\mathbb{R}^{N}$, and (iii) a readout layer that produces the output $\mathbf{y}(m)$. For an input sequence $\{u(m)\}_{m=1}^{K}$, the standard echo state network (ESN) update~\cite{Jaeger2004,Jaeger2009} reads
	\begin{subequations}
		\begin{align}
			\mathbf{x}(m) &= \mathbf{f}_{\mathrm{res}}\!\left(\mathbf{W}_{\mathrm{res}}\mathbf{x}(m-1) + \mathbf{W}_{\mathrm{in}}\,u(m)\right),
			\label{RC_eq}\\
			\mathbf{y}(m) &= \mathbf{W}_{\mathrm{out}}\,\mathbf{x}(m),
			\label{RC_out}
		\end{align}
	\end{subequations}
	where $\mathbf{f}_{\mathrm{res}}$ denotes the nonlinear activation function, $\mathbf{W}_{\mathrm{res}}\in\mathbb{R}^{N\times N}$ is the fixed recurrent connectivity matrix, and $\mathbf{W}_{\mathrm{out}}$ is the trainable readout matrix. Through the combined action of recurrence and nonlinearity, the reservoir maps the input history onto an N-dimensional dynamical state. A reservoir should satisfy the echo-state property, according to which the influence of the initial condition asymptotically vanishes and the current state is determined primarily by the recent input history. Collecting the reservoir states over time yields the state matrix $\mathbf{X}=[\mathbf{x}(1)\ \cdots\ \mathbf{x}(K)]\in\mathbb{R}^{N\times K}$, with the corresponding outputs $\mathbf{Y}=\mathbf{W}_{\mathrm{out}}\mathbf{X}$. Training minimizes the mean-squared error between the target outputs $\hat{\mathbf{Y}}$ and the predicted outputs $\mathbf{Y}$.

	\paragraph{Information processing capacity.}
	The information processing capacity is a nonlinear extension of the linear memory capacity~\cite{Yeager_shortterm}. It quantifies how well a dynamical system can reconstruct nonlinear functions of past inputs. More precisely, any time-invariant function with fading memory can be expanded using a complete orthonormal basis within a dedicated Hilbert space~\cite{dambre2012}. The components of the reservoir states in this basis are evaluated, thereby mapping the physical dynamics onto the corresponding abstract functional space. The linear memory corresponds to reconstructing delayed first-order (linear) target functions, while higher-order capacities quantify nonlinear transformations of delayed inputs. 
	
	Given the reservoir state matrix $\mathbf{X}$, we construct a family of orthogonal target functions of the past inputs using products of Legendre polynomials:
	\begin{equation}
		\hat{\mathbf{Y}}_{\{d_i\}} = \prod_{i=1}^{\mathcal{D}} \mathcal{P}_{d_i}\!\big(u(m-i)\big),
		\label{eq:polys}
	\end{equation}
	
	\noindent where $\mathcal{P}_{d_i}$ denotes the Legendre polynomial of degree $d_i$ associated with the input delayed by $i$ steps, and $\mathcal{D}$ represents the maximum delay. The corresponding capacity, obtained with an optimal linear readout, is defined as
	\begin{equation}
		C[\mathbf{X},\{d_i\}]
		= 1 -
		\frac{\min_{\mathbf{W}_{\mathrm{out}}}\left\| \hat{\mathbf{Y}}_{\{d_i\}} - \mathbf{Y}\right\|_F^2}
		{\left\| \hat{\mathbf{Y}}_{\{d_i\}} \right\|_F^2},
		\label{eq:capacity}
	\end{equation}
	where $\|\cdot\|_F$ denotes the Frobenius norm. Furthermore, after discarding the washout period, all remaining samples are used for capacity evaluation. The optimal linear readout for each target function is obtained by ordinary least-squares regression using the Moore–Penrose pseudoinverse without regularization. To prevent overestimation of small capacities arising from finite time-series length, we apply a statistical threshold of $C_{\mathrm{th}} = 0.0435$, following the procedure described in the appendix of Ref.~\cite{dambre2012} (see also Ref.~\cite{Rahul2026} for an alternative method). The total capacity, obtained by summing over an orthogonal set of polynomial target functions of the input, i.e., $C_{\text{tot}} = \sum_{\{d_i\}} C[\mathbf{X},\{d_i\}]$, is bounded by the dimension of the reservoir state space.
	
	\section{Reservoir Computing with a Synchronously Pulse-Driven Cavity}
	\label{sec:PU_RC}
	\begin{figure}[ht]
		\centering
		\begin{subfigure}[t]{0.65\textwidth}
			\vspace{0pt}
			\centering
			\includegraphics[width=\textwidth]{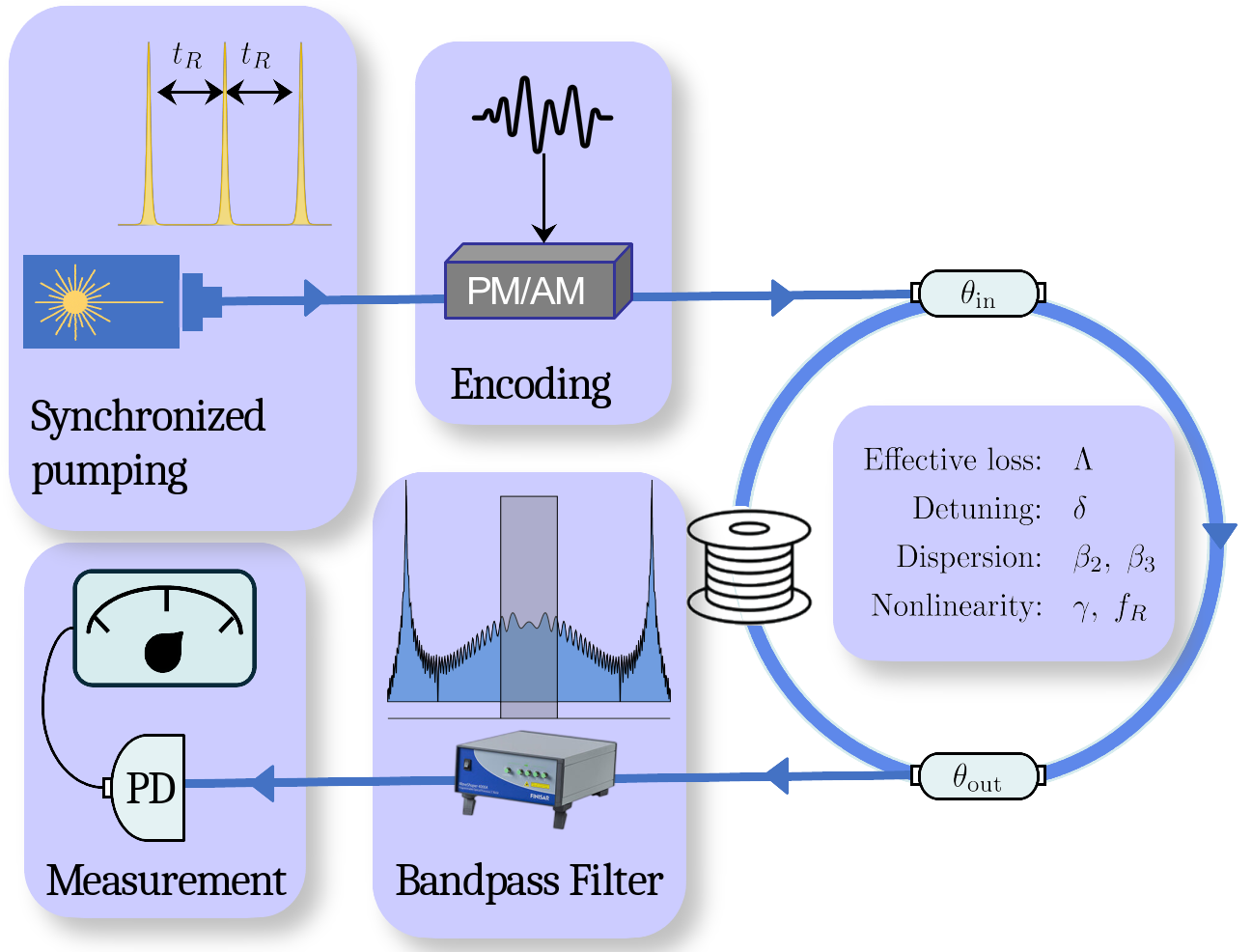}
			
			\vspace{.25cm}
			\caption{Proposed numerical architecture.}
			\label{fig:model}
		\end{subfigure}
		\hfill
		\begin{minipage}[t]{0.3\textwidth}
			\vspace{0pt}
			\begin{subfigure}[t]{\textwidth}
				\centering
				\includegraphics[width=\textwidth]{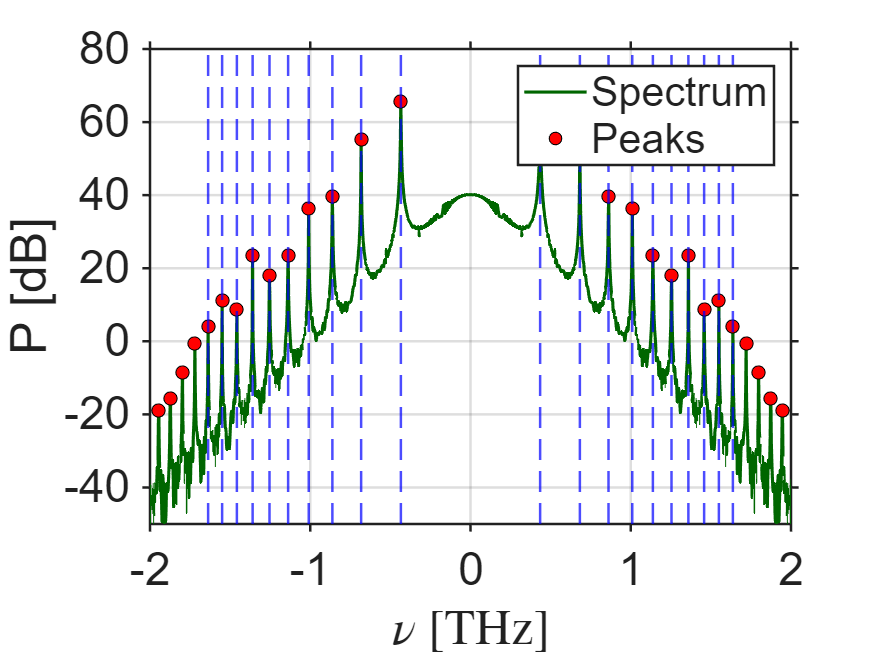}
				\caption{Spectral domain.}
				\label{fig:station-freq}
			\end{subfigure}
			\begin{subfigure}[t]{\textwidth}
				\centering
				\includegraphics[width=\textwidth]{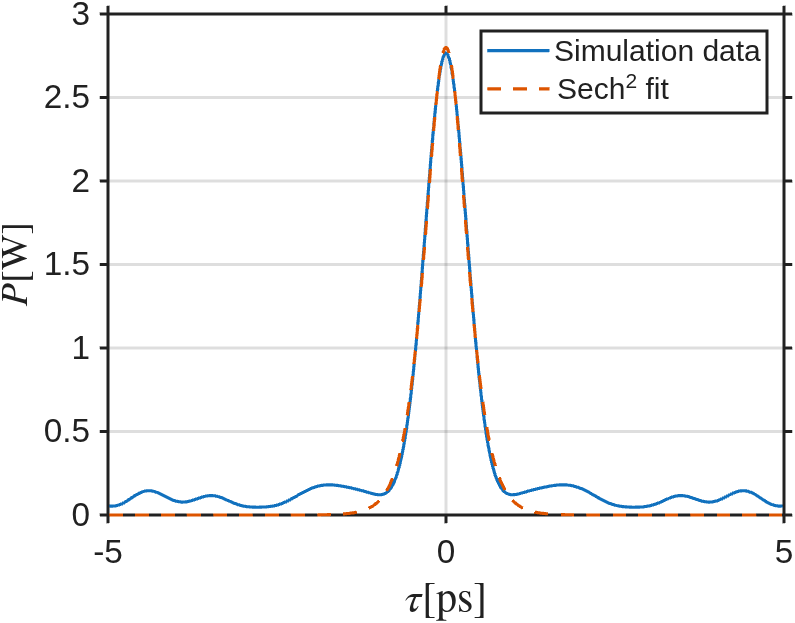}
				\caption{Temporal domain.}
				\label{fig:station-time}
			\end{subfigure}
		\end{minipage}
		
		\caption{
			Frequency-multiplexed reservoir computing with a synchronously pulse-driven optical cavity.
			\textbf{(a)} Schematic of the proposed setup. The cavity is driven by a train of $\sech$-shaped pulses whose repetition period is matched to the cavity round-trip time $t_R$. Input data are encoded onto either the phase or amplitude of the driving pulses (PM/AM). The cavity dynamics are governed by dispersion coefficients $\beta_k$, nonlinear parameters $\gamma$ and $f_R$, effective loss $\Lambda$, and input coupling ratio $\theta_{\mathrm{in}}$. The output coupler $\theta_{\mathrm{out}}$ constitutes the output layer of the reservoir and enables monitoring of the intracavity dynamics. Different nodes are accessed by applying optical bandpass filters (schematized by the gray box) to the intracavity spectrum. The spectrum shown in panel (a) is a magnified view of the spectrum in panel (b). Photodetection (PD) provides an additional quadratic nonlinearity through the conversion of optical field amplitude into optical power.
			\textbf{(b)} Steady-state intracavity spectrum. Red circles denote spectral peaks, while blue dashed lines correspond to frequencies satisfying the cavity resonance condition given by Eq.~\eqref{eq:resonantCon}.
			\textbf{(c)} Steady-state intracavity temporal profile. The blue solid line shows simulated optical power, whereas the orange dashed line corresponds to a $\mathrm{sech}^2$ fit. Panels (b) and (c) are obtained using the reference parameters listed in Table~\ref{tab:simulation_parameters}.
		}
		\label{fig:frequency_multiplexed_rc}
	\end{figure}
	The frequency-multiplexed reservoir computing architecture is schematically illustrated in Fig.~\ref{fig:model}. The cavity dynamics are modeled using the Ikeda map~\cite{ikeda1979multiple}. This model describes the propagation of the complex intracavity field in a nonlinear dispersive medium, together with a coupling boundary condition between successive roundtrips:
	\begin{subequations}
		\label{eq:ikeda_map}
		\begin{align}
			\partial_z E^{(n)}(z,\tau)
			&=
			\left(
			-\ii \frac{\beta_2}{2}\,\partial_\tau^2
			-\frac{\beta_3}{6}\,\partial_{\tau}^{3}
			+\ii\gamma \left(1 - f_R\tau_R\,\partial_{\tau}\right)\!\left|E^{(n)}(z,\tau)\right|^{2}
			\right) E^{(n)}(z,\tau),
			\label{eq:nlse}\\
			E^{(n)}(0,\tau)
			&=\sqrt{1-\Lambda}\,E^{(n-1)}(L,\tau)\,\e^{\ii\delta}
			+ \sqrt{\theta_{\mathrm{in}}}\,\mathcal{M}\!\left(E_{\mathrm{in}}(\tau), h(n)\right).
			\label{eq:bd}
		\end{align}
	\end{subequations}
	
	Equation~\eqref{eq:nlse} is known as the nonlinear Schr\"odinger equation (NLSE)~\cite{Agrawal2012} for the complex field envelope $E^{(n)}(z,\tau)$ during roundtrip $n$, where $z\in[0,L]$ is the propagation coordinate, and the fast time $\tau$ is defined in a frame co-propagating at the group velocity. Here, $\beta_j = \left.\mathrm{d}^j\beta(\omega)/\mathrm{d}\omega^j\right|_{\omega=\omega_0}$ are the dispersion coefficients evaluated at the carrier frequency $\omega_0$. Nonlinearity arises from the instantaneous Kerr effect, with coefficient $\gamma$, and delayed Raman scattering~\cite{Karpov2016,Yi2016}, where $f_R$ is the Raman fraction and $\tau_R$ denotes the first moment of the normalized Raman response in the shock approximation, valid when the relevant frequency components of the intracavity intensity lie within the approximately linear region of the Raman-gain spectrum. 
	For the first-order Raman approximation, we use \(f_R=0.18\) and \(\tau_R=16.7\,\mathrm{fs}\), corresponding to an effective Raman time \(T_R=f_R\tau_R\approx3\,\mathrm{fs}\), consistent with values reported for silica optical fibers~\cite{Atieh1999}.
	
	The boundary condition~\eqref{eq:bd} couples successive roundtrips. The parameter $\theta_{\mathrm{in}}$ denotes the input coupling ratio, while $\delta$ denotes the phase detuning between the driving laser and the nearest cavity resonance. The parameter $\Lambda$ represents the effective loss, and hence $\sqrt{1-\Lambda}$ represents the total field transmission at each roundtrip. Note that intracavity amplification can partially compensate for coupling and propagation loss~\cite{Englebert2021}, allowing the effective loss $\Lambda$ to be smaller than the input coupling ratio $\theta_{\mathrm{in}}$. We therefore treat $\Lambda$ as a lumped net-loss parameter, independent of the input coupling ratio $\theta_{\mathrm{in}}$. The modulation operator $\mathcal{M}(\cdot)$ implements either amplitude modulation, $\mathcal{M}(E_{\mathrm{in}},h(n))=(a_0 + h(n))\,E_{\mathrm{in}}$, or phase modulation, $\mathcal{M}(E_{\mathrm{in}},h(n))=E_{\mathrm{in}}\,\e^{\ii h(n)}$, where \(h(n)\) denotes the input signal applied at roundtrip \(n\). For amplitude encoding, the offset \(a_0\) is chosen after input normalization such that \(\min_n\left(a_0 + h(n)\right) = 0.05\), ensuring strictly positive driving-pulse amplitudes. For phase encoding, the positivity constraint is not applicable, and the symbols are used directly after normalization.
	
	In the anomalous-dispersion regime, and in the absence of third-order dispersion and Raman scattering, the NLSE in Eq.~\eqref{eq:nlse} admits exact soliton solutions with a hyperbolic-secant temporal envelope~\cite{Agrawal2012}. Hence, we choose the synchronized driving pulse to correspond to the fundamental soliton,
	\(
	E_{\mathrm{in}}(\tau)=\sqrt{P_0}\,
	\sech\!\left(\tau/T_0\right),
	\)
	where $P_0=61.5~\mathrm{W}$ is the unmodulated peak power and $T_0 \approx 0.54~\mathrm{ps}$ is the characteristic pulse duration, corresponding to an intensity FWHM of $0.95~\mathrm{ps}$. This choice provides a representative starting point and is further motivated by the availability of fiber-based laser sources capable of generating approximately sech$^2$-shaped pulses.
	
	Figs.~\ref{fig:station-freq} and~\ref{fig:station-time} show the spectral and temporal profiles of the stationary solution in the anomalous-dispersion regime for the reference parameters listed in Table~\ref{tab:simulation_parameters}. In the temporal domain, at a cavity detuning of $\delta=2~\mathrm{rad}$, the intracavity stationary field exhibits an approximate hyperbolic-secant envelope, corresponding to a $\sech^2$-shaped intensity profile with a peak power of $2.8~\mathrm{W}$ and an intensity FWHM of approximately $0.7~\mathrm{ps}$ (see Fig.~\ref{fig:station-time}). This intracavity stationary solution has most of its spectral power concentrated near the pump frequency, supporting the use of the first-order Raman approximation in Eq.~\eqref{eq:nlse}. The spectral peaks observed in Fig.~\ref{fig:station-freq} correspond to linear cavity modes excited by the broadband driving field and manifest as oscillatory tails in the temporal profile shown in Fig.~\ref{fig:station-time}. Numerical simulations performed in the absence of Kerr nonlinearity show that these peaks persist, confirming their linear origin, while Kerr-induced intracavity reshaping only slightly redistributes the spectral power among the resonantly enhanced components. The resonance frequencies satisfy
	\begin{equation}
		\Omega_r = \pm\sqrt{\frac{2(2\pi r-\delta)}{\beta_2L}},
		\qquad r\in\mathbb{Z},
		\label{eq:resonantCon}
	\end{equation}
	where $\Omega_r$ denotes the frequency offset of the $r$th resonance, with the corresponding resonance frequencies indicated by the blue dashed lines in Fig.~\ref{fig:station-freq}.
	
	\begin{table}[t]
		\centering
		\caption{Reference values and parameter ranges used in the simulations.}
		\label{tab:simulation_parameters}
		\setlength{\tabcolsep}{4pt}
		\renewcommand{\arraystretch}{0.95}
		\begin{tabular}{@{}lcc@{\hspace{1.5em}}lcc@{}}
			\toprule
			Parameter & Reference & Values explored
			& Parameter & Reference & Values explored \\
			\midrule
			\(\beta_2\,[\mathrm{ps^2\,km^{-1}}]\) & \(-23\) & \(\numrange{-24}{24}\) & \(L\,[\mathrm{m}]\) & 50 & 50 	\\
			\(\delta\,[\mathrm{rad}]\)	& \(2\) & \(\numrange{0}{6}\) and \(4\) &\(\gamma\,[\mathrm{W^{-1}\,km^{-1}}]\) & \(1.3\) & \(0\) \\
			\(\Lambda\,[\%]\) & \(3\) & \(\numrange{1}{10}\) &\(\tau_{\mathrm R}\,[\mathrm{fs}]\) & 0 & \(16.7\) \\
			\(\sigma_u\) & \(0.3\) & \(\numrange{0.01}{1.0}\) &\(f_{\mathrm R}\) & \(0\) & \(0.18\)\\
			\(q\) & \(30\) & \(\numrange{3}{30}\) &\(\beta_3\,[\mathrm{ps^3\,km^{-1}}]\) & \(0\) & \(0.1\) \\
			\(\theta_{\mathrm{in}}\) & \(0.1\) & \(\numrange{0.1}{0.4}\) & Modulation & Amplitude & Amplitude, phase\\
			\bottomrule
		\end{tabular}
	\end{table}
	
	\paragraph{Reservoir computing implementation.}
	The proposed system exhibits the main characteristics required for physical reservoir computing~\cite{Abreu2024}. First, its broadband intracavity spectrum provides a high-dimensional state space that can be accessed through frequency multiplexing, with each spectral channel acting as a distinct computational node. Second, nonlinear processing arises from both the Kerr and Raman dynamics described by Eq.~\eqref{eq:nlse} and the quadratic field-to-intensity transformation introduced by photodetection at the readout. Finally, the finite photon lifetime of the lossy cavity provides fading memory: the current intracavity state depends on the recent input history, whereas the influence of sufficiently old inputs and of the initial condition progressively decays over successive roundtrips. However, the proposed system does not remain stable over the entire parameter space. In the anomalous-dispersion regime, chaotic dynamics can arise [see the left panel of Fig.~\ref{fig:IPC_big}], whereas in the bistable normal-dispersion regime, the system can switch between stable states [see Fig.~\ref{fig:bistability}]. When operated in a stable input-driven regime, the reservoir response to a given input is reproducible, consistent with the practical requirements of the echo-state property.

	A total of $K=5000$ input symbols are drawn i.i.d.\ from a uniform distribution on $[0,1]$. The inputs are initially z-score normalized to have zero mean and unit standard deviation. Each discrete input $u(m)$ is held for $q$ roundtrips and scaled by a modulation depth $M$ to yield the injected signal $Mu(m)$. Consequently, the input standard deviation becomes $\sigma_u=M$, which we use as a measure of the effective input strength. Hence, the driving signal \(h(n) = Mu(m)\,|\,n\in [(m-1)q+1, mq]\).
	
	The reservoir is configured with \(N=50\) equally spaced, non-overlapping frequency channels, each with a bandwidth of \(140\,\mathrm{GHz}\), spanning a central spectral window of \(7\,\mathrm{THz}\). The value of the $i$th node at a discrete time step $m$ is defined as the average energy in that frequency channel over the $q$ roundtrips associated with input $u(m)$. For the information-processing capacity analysis, 10\% of the total inputs are discarded, and we set the maximum delay to \(\mathcal{D}=35\), which is sufficiently large to capture all non-negligible delayed contributions. The Legendre-polynomial targets are constructed after rescaling the input to the Legendre domain \([-1,1]\). To illustrate the statistical variability of the IPC estimates, error bars representing one standard deviation over ten independent input realizations are shown for the reference parameters, for both amplitude and phase encoding in Figs.~3(a) and 4(d), respectively.
	
	Unless otherwise stated, the simulations consider a standard SMF-28 fiber cavity, with the reference parameter values listed in Table~\ref{tab:simulation_parameters}. The propagation equation over each cavity roundtrip was solved using the split-step Fourier method with a fixed longitudinal step size of \(dz=0.2\,\mathrm{m}\). The fast-time window was set to \(400\,\mathrm{ps}\), corresponding to a frequency resolution of \(2.5\,\mathrm{GHz}\), and was uniformly discretized using \(2^{14}\) points. Amplitude encoding is used by default; deviations from this choice are stated explicitly. The polynomial expansion in Eq.~\eqref{eq:polys} is truncated at fourth order, as higher-order contributions are negligible in the reference configuration [see the bars labeled “Reference” in Fig.~\ref{fig:symmetry_breaking}].

	\section{Influence of Physical Parameters on Reservoir Performance}
	\label{sec:FOI}
	\begin{figure}[!ht]
		\centering
		\begin{subfigure}[t]{\textwidth}
			\centering
			\includegraphics[width=\textwidth, keepaspectratio]{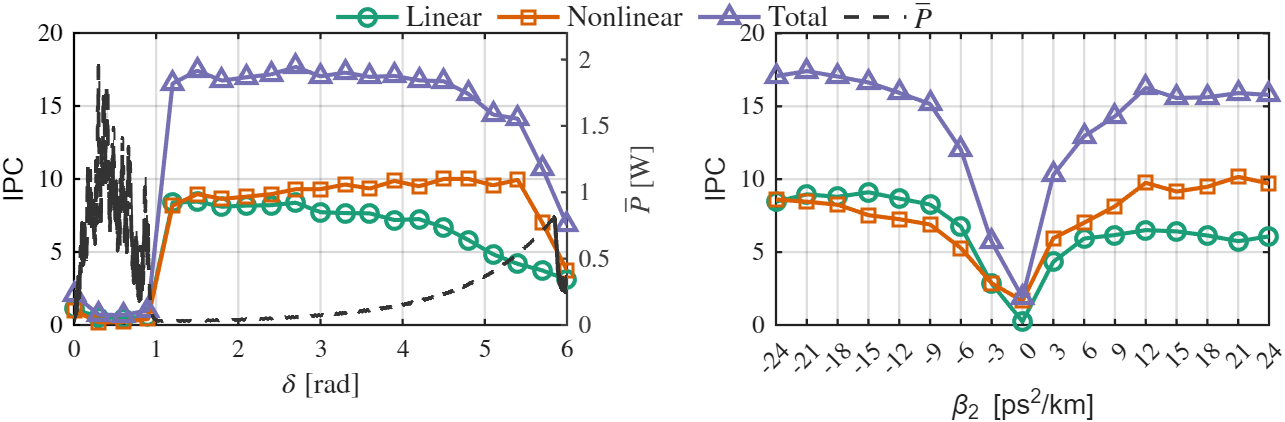}
			\caption{IPC vs detuning $\delta$ (left) and GVD $\beta_2$ (right).}
			\label{fig:IPC_big}
		\end{subfigure}
		\begin{subfigure}[c]{0.55\textwidth}
			\centering
			\includegraphics[width=\textwidth,keepaspectratio]{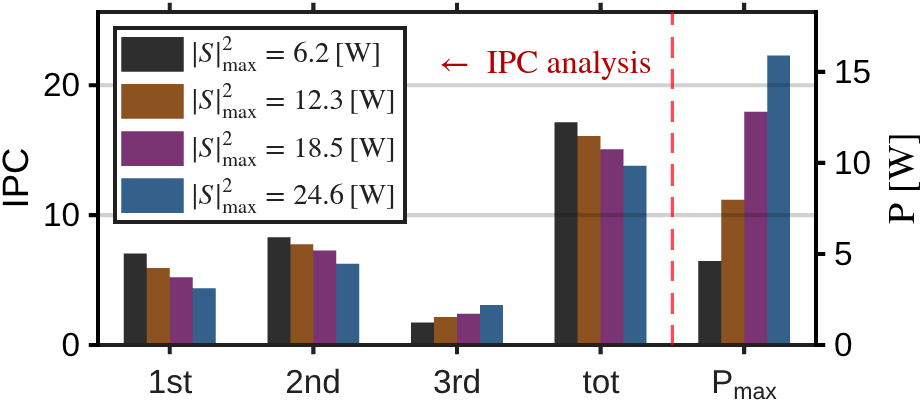}
			\caption{IPC vs driving strength $|S|^2_{\max}$.}
			\label{fig:coupler}
		\end{subfigure}
		\hspace{5pt}
		\begin{subfigure}[c]{0.4\textwidth}
			\centering
			\includegraphics[width=\textwidth,keepaspectratio]{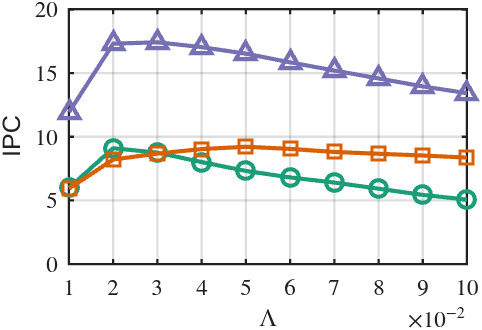}
			\caption{IPC vs effective loss $\Lambda$.}
			\label{fig:lossPhase}
		\end{subfigure}
		\caption{
			Contribution of different physical parameters to IPC, with sweep ranges specified in Table~\ref{tab:simulation_parameters}. Linear corresponds to linear memory capacities i.e., linear terms in Eq.~\eqref{eq:polys}, whereas Nonlinear represents the accumulated capacity of higher-order terms.
			\textbf{(a)} IPC as a function of cavity detuning $\delta$ (left) and group-velocity dispersion $\beta_2$ (right).
			Left: The black dashed curve represents the mean intracavity power, whose oscillations near $\delta=0$ are associated with the chaotic regime and the collapse of the information-processing capacity. Within the stationary regime, higher detuning enhances nonlinear capacity while reducing linear memory, leaving total capacity nearly unchanged.
			Right: Total IPC decreases as $|\beta_2| \to 0$; anomalous dispersion is dominated by linear memory, whereas normal dispersion promotes nonlinear processing.
			\textbf{(b)} Effect of the driving strength, $|S|_{\max}^2$, on the IPC at a cavity detuning of $\delta = 4~\mathrm{rad}$, where
			$S(\tau)=\sqrt{\theta_{\mathrm{in}}}\,E_{\mathrm{in}}(\tau)$, such that
			$|S|_{\max}^2=\theta_{\mathrm{in}}P_0$.
			Left axis: IPC as a function of the driving strength $|S|_{\max}^2$.
			Right axis: Intracavity peak power $P_{\mathrm{max}}$ corresponding to different driving strengths.
			Increasing the driving strength raises the intracavity peak power and thereby enhances the higher-order capacities.
			\textbf{(c)} IPC as a function of the effective loss $\Lambda$. The total IPC exhibits a non-monotonic dependence on effective loss, reaching a maximum near \(\Lambda=0.02\)–\(0.03\), beyond which increasing loss reduces linear memory capacity and total IPC.
		}
		\label{fig:CavityIPC}
	\end{figure}
	
	In this section, we neglect higher-order perturbative effects, namely third-order dispersion and Raman scattering, and investigate how the system performance around the reference operating regime depends on the dominant physical parameters in Eq.~\eqref{eq:ikeda_map}. Since the first- and second-order capacities dominate in the reference regime, we distinguish between the linear contribution, corresponding to the first-order capacity, and the nonlinear contribution, defined as the sum of the higher-order capacities considered in the analysis.
	
	In the anomalous-dispersion regime, the dependence of the IPC and average intracavity power on the cavity detuning is shown in the left panel of Fig.~\ref{fig:IPC_big}. The system initially exhibits chaotic dynamics at low detuning, where the information-processing capacity collapses. As the cavity detuning is increased, the field evolves toward a stable stationary state. Beyond this transition, a further increase in detuning generally leads to higher average intracavity power, accompanied by an enhancement of nonlinear IPC and a reduction in linear memory capacity.
	
	The right panel of Fig.~\ref{fig:IPC_big} illustrates the strong dependence of the IPC on the group-velocity-dispersion (GVD) coefficient $\beta_2$. In the anomalous-dispersion regime, decreasing the magnitude of $\beta_2$ reduces the nonlinear contribution to the IPC, and the reservoir response becomes increasingly linear. In both the dispersionless and normal-dispersion regimes, the system exhibits optical bistability. In the dispersionless cavity, however, the intracavity field does not converge to a stationary response; instead, the average intracavity power exhibits a modulation on a characteristic scale of approximately 20 roundtrips. These persistent internal dynamics reduce the reproducibility of the reservoir state for a given input history, thereby violating the echo-state requirement and leading to the observed collapse of the information-processing capacity. Moreover, in the normal-dispersion regime, the data correspond to the lower-branch state of the bistable cavity, reached under amplitude modulation (see Sec.~\ref{sec:bistable} for details). In this regime, the nonlinear contribution to the IPC becomes dominant.
	
	In the boundary condition of Eq.~\eqref{eq:bd}, the effective driving field is $S(\tau)=\sqrt{\theta_{\mathrm{in}}}\,E_{\mathrm{in}}(\tau)$. Hence, the peak driving power, $|S|_{\max}^2=\theta_{\mathrm{in}}P_0$, can be varied either through the input coupling ratio or through the peak power of the driving pulse. Here, we keep the pulse profile and $P_0$ fixed and vary $\theta_{\mathrm{in}}$ as a convenient means of controlling the driving strength. We numerically observe that the chaotic regime expands with increasing $|S|_{\max}^2$. We stabilize the cavity at the higher detuning $\delta=4~\mathrm{rad}$. Figure~\ref{fig:coupler} shows that increasing the driving strength $|S|_{\max}^2$ raises the intracavity peak power, which consequently increases the contributions of third-order polynomials. Notably, the apparent reduction in total capacity at higher values of $|S|_{\max}^2$ is consistent with a redistribution of information-processing capacity toward higher-order polynomial terms that are difficult to estimate and are not fully captured by the truncated expansion.
	
	Figure~\ref{fig:lossPhase} shows a non-monotonic dependence of the IPC on the effective roundtrip loss. This behavior can be interpreted in terms of the competition between the cavity photon lifetime, $t_{\mathrm{ph}}\simeq t_R/\Lambda$, and the symbol duration associated with the repetition factor, $T_{\mathrm{sym}}=q t_R$. Accordingly, the fraction of intracavity power retained after one input window scales approximately as $\exp(-q\Lambda)$. For $q\Lambda\ll1$, the symbol duration is much shorter than the cavity lifetime, such that the intracavity field changes only weakly before the next input symbol is applied. This slow response accounts for the reduced IPC observed at the smallest $\Lambda$. Conversely, for $q\Lambda\gg1$, the cavity substantially relaxes within each input window, reducing the memory retained between successive symbols and therefore decreasing the linear memory capacity. An intermediate regime, $q\Lambda\sim1$, provides a balance between cavity response and memory retention. For $q=30$, as used in Fig.~\ref{fig:lossPhase}, the maximum total IPC occurs around $\Lambda=0.02$--$0.03$, corresponding to $q\Lambda\simeq0.6$--$0.9$, consistent with this crossover regime.
	
	\section{Spectral Symmetry Breaking and Reservoir Performance}
	\label{sec:HOI}
	\begin{figure}[htbp]
		\centering
		\begin{subfigure}[t]{0.3\textwidth}
			\centering
			\includegraphics[width=\textwidth, keepaspectratio]{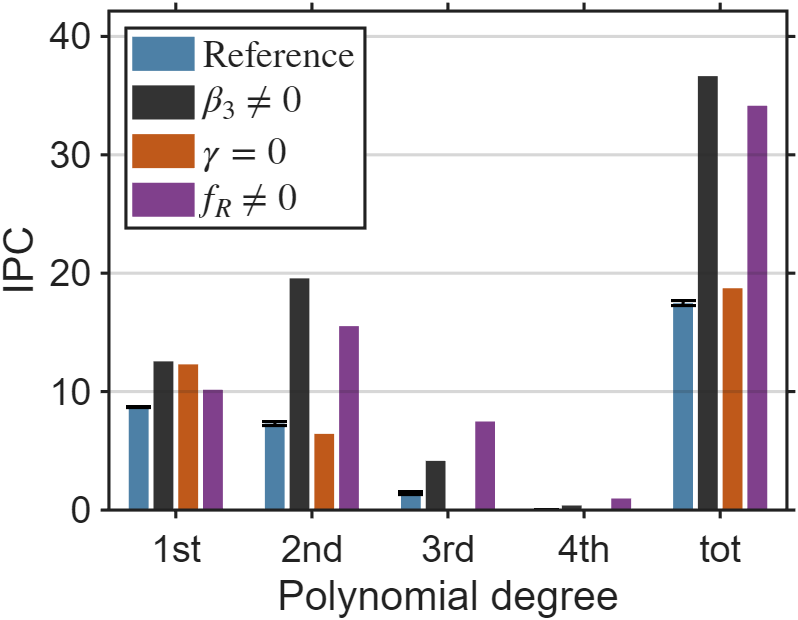}
			\caption{Symmetry-enhancement IPC.}
			\label{fig:symmetry_breaking}
		\end{subfigure}
		\hfill
		\begin{subfigure}[t]{0.3\textwidth}
			\centering
			\includegraphics[width=\textwidth, keepaspectratio]{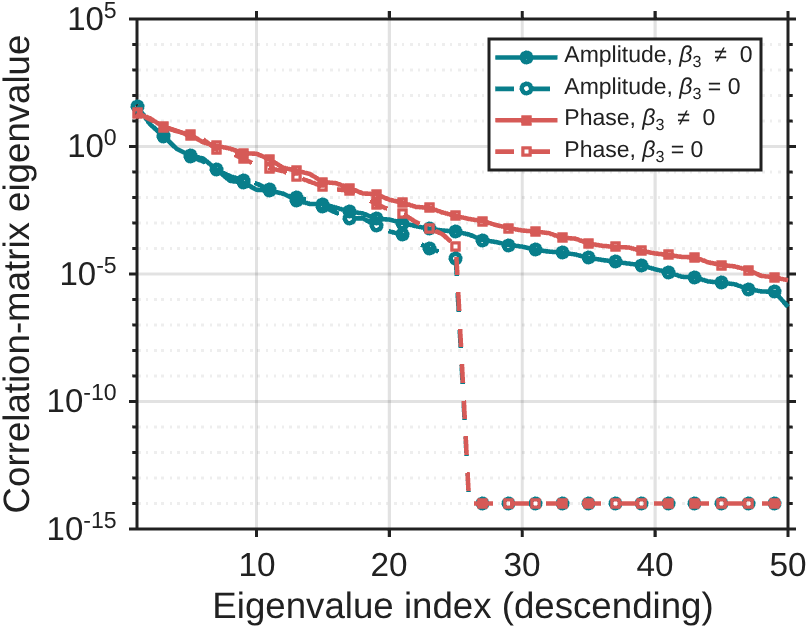}
			\caption{Eigenvalue decomposition.}
			\label{fig:eg}
		\end{subfigure}
		\hfill
		\begin{subfigure}[t]{0.3\textwidth}
			\centering
			\includegraphics[width=\textwidth, keepaspectratio]{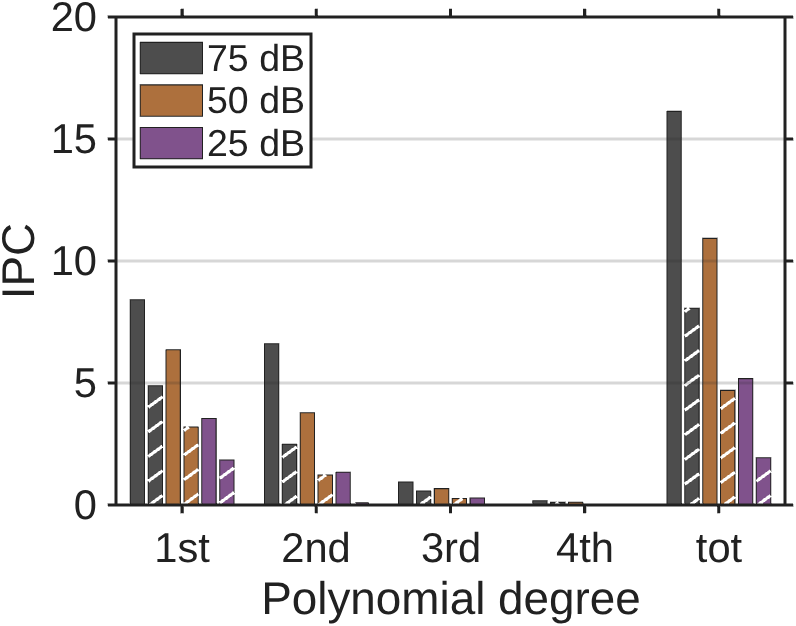}
			\caption{Impact of noise on IPC.}
			\label{fig:noise}
		\end{subfigure}
		\par\vspace{5pt}
		\begin{subfigure}[t]{\textwidth}
			\centering
			\includegraphics[width=\textwidth, keepaspectratio]{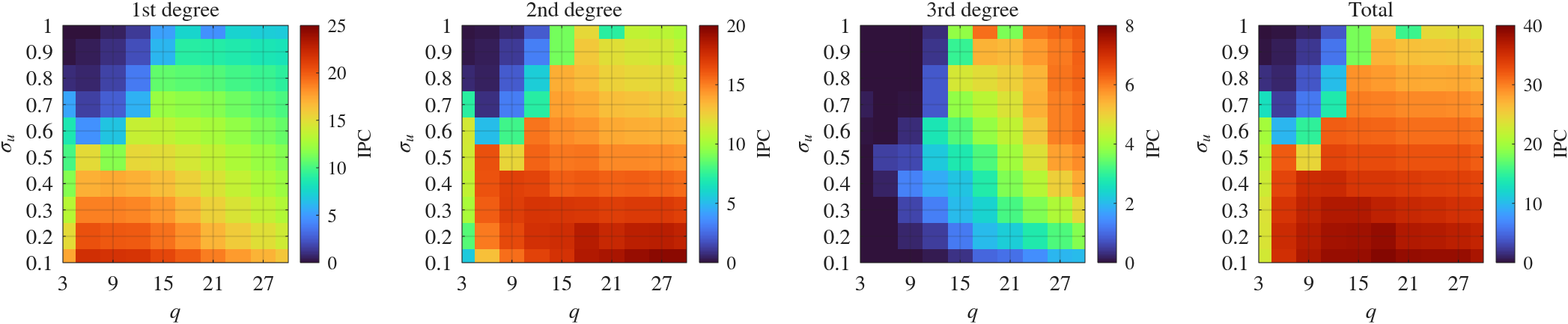}
			\caption{IPC landscape for phase-encoded inputs in the $(q,\sigma_u)$ parameter space.}
			\label{fig:encodings}
		\end{subfigure}
		\par\vspace{5pt}
		\begin{subfigure}[t]{\textwidth}
			\centering
			\includegraphics[width=\textwidth, keepaspectratio]{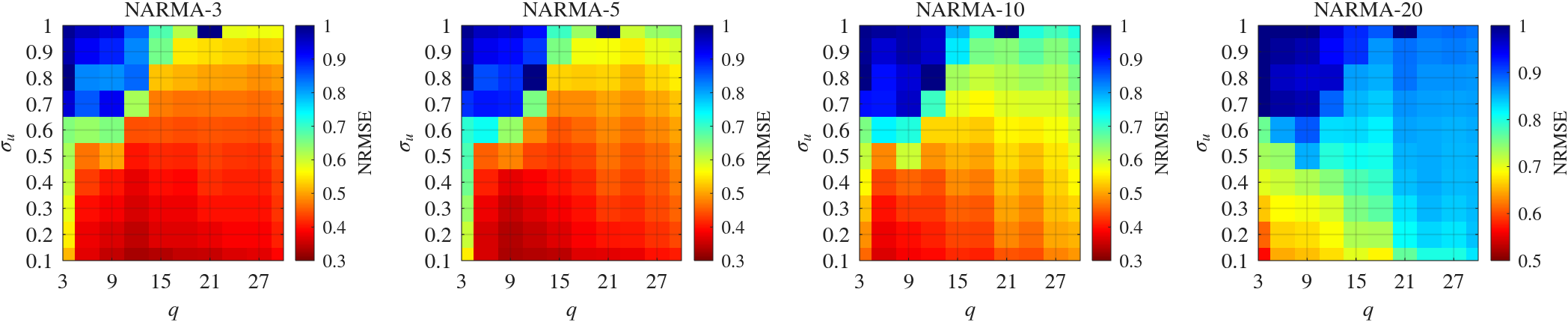}
			\caption{NRMSE of the NARMA benchmarks in the $(q,\sigma_u)$ parameter space.}
			\label{fig:narma}
		\end{subfigure}
		\caption{
			Impact of spectral symmetry breaking, noise, and phase encoding on reservoir information processing. Panels~(c)--(e) correspond to the configuration with $\beta_3 \neq 0$ and $f_R = 0$.
			\textbf{(a)} IPC comparison between the reference case and three physical perturbations: third-order dispersion ($\beta_3 \neq 0$), suppressed Kerr nonlinearity ($\gamma = 0$), and Raman scattering ($f_R \neq 0$). Third-order dispersion and Raman scattering break spectral symmetry around the pump frequency, reducing redundancy between opposite frequency channels and enhancing total IPC. Suppressing Kerr nonlinearity reduces higher-order nonlinear memory contributions. Error bars illustrate the statistical variability of the IPC estimates for the reference configuration, representing one standard deviation over ten independent input realizations.
			\textbf{(b)} Eigenvalue spectra of the reservoir-state correlation matrices for amplitude- and phase-encoded inputs, with (\(\beta_3\neq0)\) and without (\(\beta_3=0)\) third-order dispersion. In the absence of third-order dispersion, half of the eigenvalues vanish up to numerical precision due to redundancy between mirror-symmetric spectral channels. Introducing third-order dispersion lifts this degeneracy, increasing the effective dimensionality of the reservoir.
			\textbf{(c)} Impact of noise on reservoir-computing performance. Solid bars indicate amplified spontaneous emission (ASE) noise, while single-hatched bars indicate output detection noise. Different bar colors correspond to different signal-to-noise ratios (SNRs).
			\textbf{(d)} IPC decomposition for phase-encoded inputs as functions of the input standard deviation $\sigma_u$ and repetition factor $q$. Increasing $\sigma_u$ and $q$ redistributes the information-processing capacity from lower- to higher-order polynomial contributions in Eq.~\eqref{eq:polys}.
			\textbf{(e)} NRMSE of the NARMA-$R$ benchmarks in the $(q,\sigma_u)$ parameter space for $R=\{3,5,10,20\}$. The minimum NRMSE is obtained at relatively low input modulation strengths $\sigma_u$. Increasing the NARMA order increases the temporal-processing complexity of the task and results in progressively larger NRMSE values over the investigated parameter range.
		}
		\label{fig:IPC_enhance}
	\end{figure}
	The performance of the system can be further improved either by reducing redundancy between mirror-symmetric frequency channels or by modifying the information-encoding scheme. In the absence of third-order dispersion (\(\beta_3=0)\) and Raman scattering (\(f_R=0)\), Eq.~\eqref{eq:nlse}, together with a symmetric driving pulse, is invariant under temporal reflection $\tau\rightarrow-\tau$, implying an invariance of the spectrum $\vert \tilde E(0,\Omega) \vert^2$ under $\Omega \to - \Omega$ where $\Omega$ is the angular frequency offset from the pump. The  spectral symmetry produces redundancy between reservoir nodes located at opposite frequency offsets from the pump. This redundancy effectively reduces the dimensionality of the reservoir and can be mitigated by breaking the symmetry of the intracavity dynamics. 
	
	Third-order dispersion breaks the spectral symmetry through its odd-order frequency-dependent phase contribution, whereas Raman scattering introduces a delayed nonlinear response that leads to asymmetric spectral energy transfer. This, in turn, diversifies the dynamics of equally spaced spectral channels, thereby increasing the system's IPC as shown in Fig.~\ref{fig:IPC_enhance}. To provide quantitative evidence for the reduction of spectral-node redundancy induced by symmetry breaking, we investigate the eigenvalue spectrum of the reservoir-state correlation matrix, focusing on the effect of third-order dispersion. Since amplitude and phase modulation perturb the driving field in physically distinct ways, we consider both encoding schemes, with and without third-order dispersion. We construct the Pearson correlation matrix of the reservoir states and calculate its eigenvalues, arranged in descending order. As shown in Fig.~\ref{fig:eg}, in the absence of third-order dispersion (\(\beta_3=0\)), half of the eigenvalues vanish up to numerical precision, reflecting the redundancy between mirror-symmetric spectral nodes and the corresponding reduction in the effective dimensionality of the reservoir by a factor of two. Introducing third-order dispersion lifts this degeneracy and produces a more gradual decay of the eigenvalue spectrum, indicating the emergence of additional independent directions in the reservoir-state space. To further quantify this effect, we evaluate the mean absolute Pearson correlation coefficient between mirror-symmetric spectral nodes. In the absence of third-order dispersion, the mean mirror-node correlation is unity for both encoding schemes. When third-order dispersion is introduced, this correlation decreases to approximately 0.4 for phase encoding and 0.8 for amplitude encoding.
	
	Noise is a major limiting factor in analog computing. Figure~\ref{fig:noise} shows how different noise sources affect the IPC of the system. We consider the configuration including third-order dispersion, for which the total IPC is approximately $36$. One practical route to realizing the low effective loss considered here is to use an active fiber cavity in which an intracavity gain medium, operated below the lasing threshold, partially compensates the roundtrip losses~\cite{Englebert2021}. Such gain inevitably introduces amplified spontaneous emission (ASE) into the cavity, which can perturb the reservoir dynamics. For a given signal power \(P_{\mathrm{sig}}\), the noise power is set according to \(P_{\mathrm{noise}} = P_{\mathrm{sig}}/\mathrm{SNR}_{\mathrm{lin}}\), where \(\mathrm{SNR}_{\mathrm{lin}}\) denotes the prescribed linear signal-to-noise ratio. The SNR values reported in Fig.~\ref{fig:noise} are expressed in decibels according to \(\mathrm{SNR}_{\mathrm{dB}} = 10\log_{10}(\mathrm{SNR}_{\mathrm{lin}})\). We separately investigate two noise sources relevant to experimental implementations: intracavity ASE noise (solid bars), modeled as additive complex Gaussian noise injected into the intracavity field at every roundtrip, and detection-stage noise (single-hatched bars), for which noise is added to the reservoir-state matrix. Both noise sources substantially reduce the reservoir's IPC. Noise obscures the dynamics of low-power spectral channels, particularly those at large frequency offsets from the pump, thereby reducing the effective number of computational nodes.
	
	Different machine-learning tasks require different balances between linear memory and nonlinear processing, so controlling the reservoir response is an important aspect of system optimization. We tune this balance by varying \(\sigma_u\) and \(q\). For amplitude encoding, in the absence of Kerr nonlinearity (\(\gamma=0\)), the third- and higher-order nonlinear memory capacities vanish, while the remaining nonlinear processing arises solely from the square-law transformation introduced by photodetection; accessing higher-order capacities therefore relies on the Kerr nonlinearity of the cavity. Amplitude encoding is constrained by the transmission range of a passive modulator. For a fixed nominal source field, the amplitude-transmission coefficient must satisfy \(a_0+h(n)\leq1\). However, when this condition is exceeded, the same injected field can be realized by increasing the source-field amplitude and rescaling the transmission coefficients. In the present study, we restrict the amplitude-encoding scan to \(\sigma_u\leq0.3\). 
	 
	 By contrast, phase encoding keeps the driving-pulse amplitude constant and, as shown in Fig.~\ref{fig:encodings}, maintains stable operation over a broader range of \(\sigma_u\) and \(q\). As a result, higher-order polynomial capacities become accessible at larger modulation depths and larger values of \(q\). At	small \(q\) and large \(\sigma_u\), however, the low-performance region of Fig.~\ref{fig:encodings} (and Fig.~\ref{fig:narma}) is consistent with the combined effect of insufficient cavity response within one symbol interval (\(q\Lambda\ll1\)) and strong modulation, which can drive the system away from the stable input-driven operating regime.
		
	\paragraph{NARMA benchmark.}
	The nonlinear autoregressive moving-average (NARMA) time series is a widely used benchmark for evaluating recurrent neural networks and reservoir-computing systems~\cite{Atiya2000}. It requires the simultaneous processing of nonlinear transformations and delayed information. For a NARMA process of order $R$, the target sequence $d(n)$ is generated recursively according to
	\begin{equation}
		d(n+1)=0.3d(n)+0.05d(n)
		\sum_{i=0}^{R-1} d(n-i)+1.5u(n-R+1)u(n)+0.1.
		\label{eq:narma}
	\end{equation}
	
	The input sequence $u(n)$ is drawn i.i.d.\ from a uniform distribution on $[0,0.2]$ and is used to generate the NARMA target sequence $d(n)$ according to Eq.~\eqref{eq:narma}. The reduced input range is chosen to mitigate divergence of the target dynamics at higher NARMA orders. For injection into the reservoir, the same input sequence is z-score normalized and rescaled to the prescribed input standard deviation $\sigma_u$, following the preprocessing procedure described in Sec.~\ref{sec:PU_RC}. We consider NARMA orders $R=\{3,5,10,20\}$, thereby progressively increasing the temporal-processing complexity of the benchmark. After discarding the first $10\%$ of the samples as a washout period, the remaining time series is divided into training and test sets. The first $70\%$ of the samples, corresponding to the reservoir states $X_{\mathrm{train}}$ and target sequence $\hat{Y}_{\mathrm{train}}$, are used to train the linear readout. The remaining $30\%$ are used exclusively for testing. The readout weights are obtained by ordinary least-squares regression, without regularization, and the prediction performance is quantified using the normalized root-mean-square error (NRMSE).
	 
	 Figure~\ref{fig:narma} shows the reservoir performance for different NARMA orders $R$ as a function of $(\sigma_u,q)$. The lowest NRMSE values are generally obtained for relatively small modulation strengths $\sigma_u$, where the intracavity response remains close to a linear mapping, while the square-law photodetection provides a quadratic transformation of the optical field. In this regime, substantial first- and second-order information-processing capacities are retained. This behavior is consistent with Eq.~\eqref{eq:narma}, whose elementary nonlinear terms are quadratic. Since the broad parameter scan shown in Fig.~\ref{fig:narma} begins at $\sigma_u=0.1$, we performed an additional fine scan over $\sigma_u=0.01$--$0.09$ in steps of $0.02$. For all considered NARMA orders, this refinement yields the minimum NRMSE at $\sigma_u=0.03$. The corresponding minimum NRMSE values are $0.302$, $0.318$, $0.353$, and $0.545$, respectively, obtained at repetition factors $q=15$, $12$, $6$, and $3$. For NARMA-$10$, the minimum NRMSE of $0.353$ is comparable to the experimental result of Ref.~\cite{Vinckier2015}, which reported an NMSE of $0.107\pm0.012$, corresponding to an NRMSE of approximately $0.327\pm0.018$.
	
	\section{Reservoir Computing in the Optically Bistable Regime}
	\label{sec:bistable}
	\begin{figure}[!ht]
		\centering
		\begin{subfigure}[t]{0.9\textwidth}
			\centering
			\includegraphics[width=\textwidth,keepaspectratio]
			{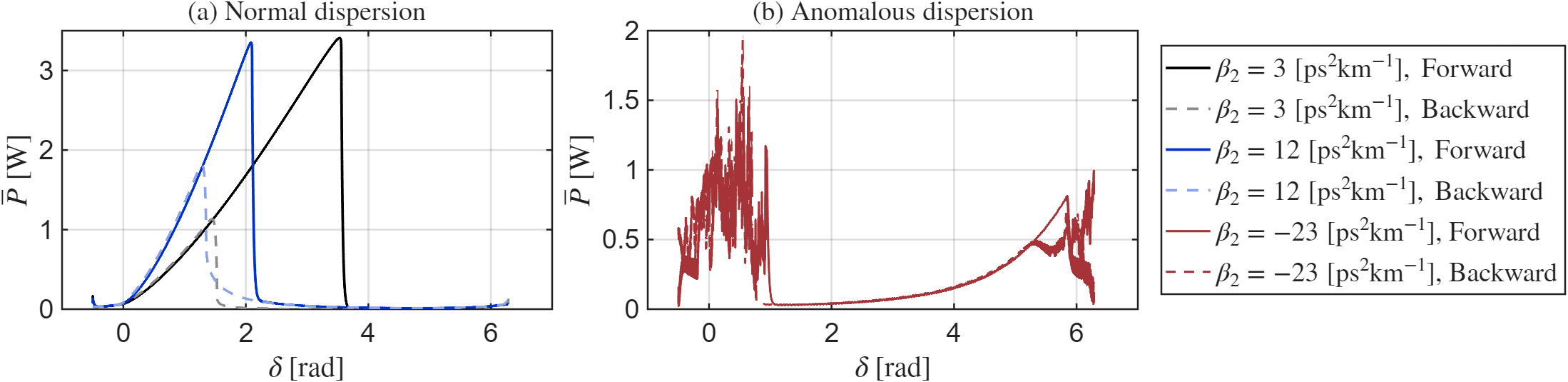}
			\caption{Bistability under normal and (b) anomalous dispersion.}
			\label{fig:normal1}
		\end{subfigure}
		\setcounter{subfigure}{2}
		\begin{subfigure}[t]{0.55\textwidth}
			\centering
			\includegraphics[width=\textwidth,keepaspectratio]
			{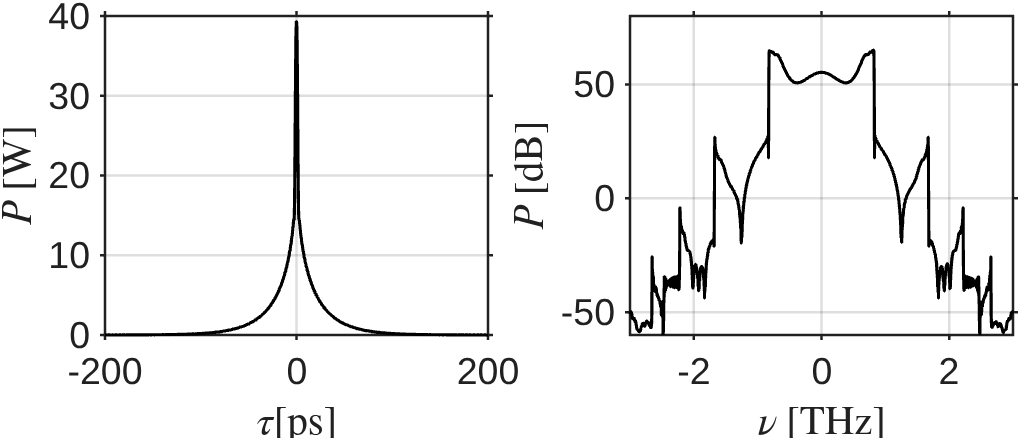}
			\caption{Upper-branch profiles for
				$\beta_2=3~\mathrm{ps^2\,km^{-1}}$.}
			\label{fig:normal2}
		\end{subfigure}
		\hspace{10pt}
		\begin{subfigure}[t]{0.30\textwidth}
			\centering
			\includegraphics[width=\textwidth]
			{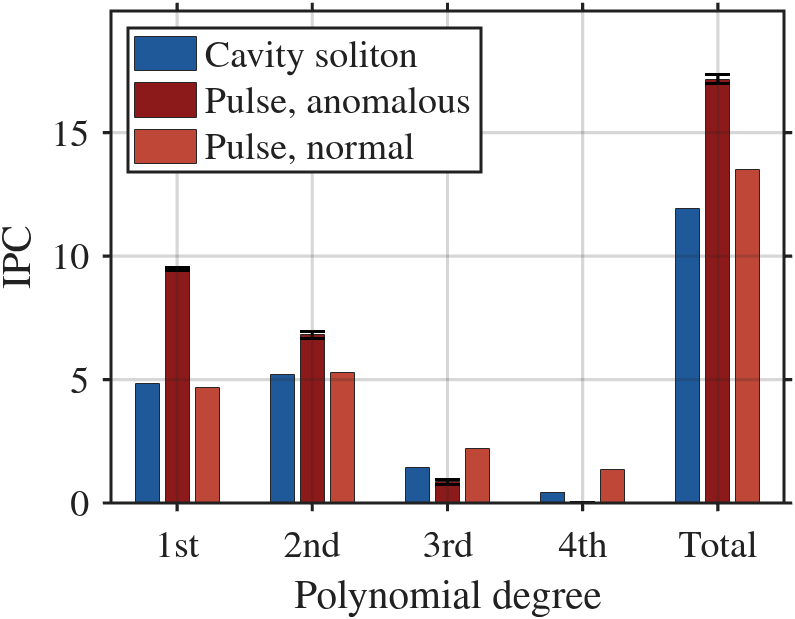}
			\caption{IPC comparison.}
			\label{fig:IPCCompare}
			\end{subfigure}
		\caption{ 
			\textbf{(a)-(b)} Synchronously pulse-driven cavity in the normal and anomalous-dispersion regime. Intracavity curves for $\beta_2 = 3~\mathrm{ps^2\,km^{-1}}$ (black) and $\beta_2 = 12~\mathrm{ps^2\,km^{-1}}$ (blue), and $\beta_2 = -23~\mathrm{ps^2\,km^{-1}}$ (red). Solid curves correspond to the forward detuning scan, while dashed curves indicate the backward scan. In the normal dispersion regime the branch termination points depend on the group-velocity dispersion.
			\textbf{(c)} Temporal and spectral profiles of the upper-branch solution of the synchronously pulse-driven cavity for $\beta_2 = 3~\mathrm{ps^2\,km^{-1}}$.
			\textbf{(d)} IPC comparison of different operating regimes of the driven cavity under phase encoding. The synchronously pulse-driven cavity in the anomalous-dispersion regime achieves the largest total IPC, while the single-cavity-soliton state and the synchronously pulse-driven cavity in the normal-dispersion regime exhibit relatively stronger higher-order contributions due to their high-peak-power intracavity states. Error bars are shown for the synchronously pulse-driven anomalous-dispersion configuration and represent one standard deviation over ten independent input realizations.}
		\label{fig:bistability}
	\end{figure}
	
	In this section, we investigate bistable operating regimes and their implications for the proposed frequency-multiplexed reservoir computer. Optical bistability occurs when the intracavity field admits multiple stable solutions for the same set of system parameters. Optical bistability and modulation instability under synchronous pulsed pumping have been investigated in Kerr resonators. In fiber Fabry–Pérot cavities, previous studies have considered quasi-CW pumping~\cite{Ziani2024,Bunel2023} and Gaussian pulses~\cite{Xu2025}, while ultrashort-pulse-driven fiber ring cavities have been investigated in both dispersion regimes~\cite{Luo2015}.
	
	To identify the corresponding bistable regions, we monitor the intracavity power while sweeping the cavity detuning in both the forward and backward directions. Figures~\ref{fig:bistability}(a) and (b) compare forward and backward detuning scans of the present pulse-driven system in the normal- and anomalous-dispersion regimes, respectively. Under normal dispersion, pronounced hysteresis loops are observed, whose branch termination points depend on the group-velocity dispersion, with smaller values of \(\beta_2\) allowing the bistable branch to persist over a wider detuning range. In the anomalous-dispersion regime, chaotic dynamics arise at small cavity detuning, leading to a collapse of the information-processing capacity and preventing reliable reservoir operation under these conditions. However, no evidence of optical bistability is observed over a broad detuning range, including the reference operating point at \(\delta=2~\mathrm{rad}\). Figure~\ref{fig:normal2} shows the temporal and spectral profiles of the upper-branch stationary solution obtained for a synchronously pulse-driven cavity with \(\beta_2=3~\mathrm{ps^2\,km^{-1}}\). Together with the anomalous-dispersion stationary profile shown in Fig.~\ref{fig:station-time}, these results illustrate the intracavity solutions, whose profiles are shaped by the combined effects of group-velocity dispersion, Kerr nonlinearity, and synchronous pulsed driving. Unlike the quasi-CW pumping considered in Ref.~\cite{Ziani2024}, these solutions cannot be described by a dispersion-independent formalism.
	
	CW-driven Kerr cavities, in which the driving term in Eq.~\eqref{eq:bd} is replaced by the constant \(\sqrt{\theta_{\mathrm{in}}P_{\mathrm{in}}}\), have been extensively studied and are known to exhibit optical bistability~\cite{Haelterman1992,lugiato1987spatial}. In the anomalous-dispersion regime, depending on the cavity detuning and driving conditions, the system may support a homogeneous CW state, modulation-instability patterns, and dissipative cavity solitons~\cite{coen2013}. Cavity solitons (CSs) are stable attractors of the system, and the intracavity field can converge to a single-CS state when initialized sufficiently close to the CS solution. For $P_{\mathrm{in}}=\SI{250}{\milli\watt}$ and a cavity detuning $\delta=2~\mathrm{rad}$, with all other parameters set to the reference values in Table~\ref{tab:simulation_parameters}, initializing the intracavity field with the pulse profile $E_{\mathrm{in}}$ introduced in Sec.~\ref{sec:PU_RC} leads to convergence toward a single-CS state comparable to that investigated in Ref.~\cite{arsalan2026} (see Refs.~\cite{coen2013,Wabnitz1993} for more details).
	
	RC implementation within a bistable system depends sensitively on the encoding scheme. For upper-branch reservoir operation, the state must remain within that branch’s basin of attraction; otherwise, sufficiently strong modulation can drive the system beyond the stability boundary and cause a transition to the lower-power state. We find that amplitude encoding is particularly disruptive, as even modest amplitude variations can induce abrupt changes in the balance between coherent driving and effective loss, causing the system to leave the upper branch and collapse onto the lower branch. Phase encoding provides a substantially larger stability margin. In our simulations with phase-encoded inputs, the single-CS state remains stable for input standard deviations up to approximately \(\sigma_u=0.4\), whereas the synchronously pulse-driven system in the normal-dispersion regime with \(\beta_2=3~\mathrm{ps^2\,km^{-1}}\) remains stable up to approximately \(\sigma_u=0.75\). We therefore adopt phase encoding with \(\sigma_u=0.3\) for all subsequent comparisons, while keeping the remaining system parameters fixed as specified in Table~\ref{tab:simulation_parameters}.
	
	The IPC results for the three systems are summarized in Fig.~\ref{fig:IPCCompare}. The largest total capacity is obtained for synchronous pulse pumping in the anomalous-dispersion regime. This enhancement is primarily due to the first-order polynomial contribution, consistent with the predominantly linear response of the intracavity field at the relatively low modulation strength considered here (see Fig.~\ref{fig:encodings} for enhanced nonlinear processing). By contrast, the higher intracavity peak power reached on the upper branch of the bistable regime produces a modest enhancement of the third- and fourth-order polynomial capacities.
	 
	\section{Conclusion}
	\label{sec:conclusion}
	
	This study complements our previous investigation of cavity-soliton-based reservoir computing by exploring a distinct synchronously pulse-driven operating regime. In both systems, we identified linear waves as a key ingredient for enhancing information-processing performance, although the underlying mechanisms differ substantially. In Ref.~\cite{arsalan2026}, cavity solitons provided the nonlinear computational substrate, while the associated Kelly sidebands~\cite{kelly1,kelly2} increased the effective dimensionality of the reservoir. By contrast, the synchronously pulse-driven cavity in the anomalous-dispersion regime is weakly nonlinear, and information processing relies primarily on the excitation of linear cavity modes. Moreover, the proposed platform can operate in both the anomalous- and normal-dispersion regimes. The corresponding operating states exhibit markedly different IPC characteristics, demonstrating that the balance between linear memory and nonlinear transformation can be controlled through group-velocity dispersion.
	
	The spectral symmetry of the intracavity field produces redundancy between reservoir nodes located at opposite frequency offsets from the pump. In frequency-multiplexed architectures, this redundancy reduces the effective dimensionality of the reservoir. Similar node redundancy has also been identified in previous frequency-multiplexed reservoir implementations \cite{arsalan2026,Cuevas2025}. In Ref.~\cite{arsalan2026}, we mitigated this redundancy at the readout stage by using an asymmetric distribution of spectral filters across the cavity-soliton spectrum. In contrast, the present work shows that the redundancy can be reduced directly through the underlying intracavity dynamics, even when the reservoir nodes are sampled on a symmetric frequency grid. In particular, higher-order dispersion and Raman scattering break the spectral symmetry and diversify the dynamics of otherwise equivalent frequency channels. The spectral asymmetry can also arise through spontaneous temporal symmetry breaking in synchronously pulse-driven Kerr resonators operating within specific driving and detuning regimes, as demonstrated experimentally in Ref.~\cite{Xu2014}. These physical mechanisms could therefore be exploited to optimize the reservoir's information-processing performance.
	
	We also assessed the robustness of the reservoir against intracavity and detection-stage noise, both of which degrade its information-processing performance. Beyond these noise sources, timing jitter and pump--cavity desynchronization represent important experimental considerations that were not included in the present analysis. Timing jitter can disrupt the coherent accumulation of the resonantly enhanced linear cavity modes, while pump--cavity desynchronization is itself an important dynamical control parameter in pulse-driven Kerr resonators and can substantially modify the stationary states, spectral characteristics, and temporal drift of the intracavity field~\cite{Hendry2019, Anderson2023}. A systematic investigation of these effects in the present reservoir-computing configuration is left for future work.
	
	Finally, we considered the normal-dispersion regime, where the synchronously pulse-driven cavity with positive Kerr nonlinearity exhibits a bistable response under short-pulse pumping. The IPC evaluation of the upper branch demonstrates that its higher intracavity peak power leads to enhanced nonlinear processing. However, operation in the bistable regime requires additional care, since sufficiently strong input modulation or other perturbations can induce switching from the upper to the lower branch. In this respect, the synchronously pulse-driven cavity in the anomalous-dispersion regime with phase-encoded inputs provides an attractive alternative: as shown in Fig.~\ref{fig:encodings}, the reservoir response can be tuned from predominantly linear to increasingly nonlinear processing by varying the modulation strength, without relying on operation on a bistable branch.

\begin{backmatter}
	\bmsection{Funding}
	We acknowledge funding through projects CDR J.0143.24, EQP U.N051.24, EOS O.0019.22 (Photonic Ising Machines), and EOS No. 40007560 (PULSE), and the ERC project HIGHRES (grant agreement No. 101125625).
	
	\bmsection{Disclosures}
	The authors declare no conflicts of interest.
	
	\bmsection{Data availability}
	Data underlying the results presented in this paper are not publicly available at this time but may be obtained from the authors upon reasonable request.
\end{backmatter}

\end{document}